# Big Data, Data Science, and Civil Rights

Solon Barocas, Elizabeth Bradley, Vasant Honavar, and Foster Provost

Abstract

Advances in data analytics bring with them civil rights implications. Data-driven and algorithmic decision making increasingly determine how businesses target advertisements to consumers, how police departments monitor individuals or groups, how banks decide who gets a loan and who does not, how employers hire, how colleges and universities make admissions and financial aid decisions, and much more. As data-driven decisions increasingly affect every corner of our lives, there is an urgent need to ensure they do not become instruments of discrimination, barriers to equality, threats to social justice, and sources of unfairness. In this paper, we argue for a concrete research agenda aimed at addressing these concerns, comprising five areas of emphasis: (i) Determining if models and modeling procedures exhibit objectionable bias; (ii) Building awareness of fairness into machine learning methods; (iii) Improving the transparency and control of data- and model-driven decision making; (iv) Looking beyond the algorithm(s) for sources of bias and unfairness—in the myriad human decisions made during the problem formulation and modeling process; and (v) Supporting the cross-disciplinary scholarship necessary to do all of that well.

Over the past several years, government, academia, and the private sector have increasingly recognized that the use of big data and data science in more and more decisions has important implications for civil rights, from racial discrimination to income equality to social justice. We have seen many fruitful meetings and discussions, some of which are summarized briefly in an appendix below and have informed this report. However, a coherent research agenda for addressing these topics is only beginning to emerge.

The need for such an agenda is critical and timely. Big data and data science have begun to profoundly affect decision making because the modern world is more broadly instrumented to gather data—from financial transactions, mobile phone calls, web and app interactions, emails, chats, Facebook posts, Tweets, cars, Fitbits, and on and on. Increasingly sophisticated algorithms can extract patterns from that data, enabling important advances in science, medicine, and commerce. As described in a recent *60 Minutes* segment, for instance, IBM's Watson has helped doctors identify treatment strategies for cancer.[1] Xerox now cedes hiring decisions for its 48,700 call-center jobs to software, cutting attrition by a fifth.[2] And if you use the web, you have received advertisements targeted based on fine-grained details of your online behavior.

Along with improved science and commerce come important civil rights implications. For example, data analytics tools can capture and instantiate decision-making patterns that are implicitly discriminatory— and can do so unintentionally, simply from distilling the data. Implicit discrimination by algorithms requires our attention because such data-driven methods are deployed in many of our most crucial social institutions. Risk assessment tools, for instance, are increasingly common in the criminal justice system, informing critical decisions like pre-trial detention, bond amounts, sentence lengths, and parole. Last year, *ProPublica* completed a study of a risk assessment tool employed in a number of courtrooms across that nation that is equally accurate in predicting whether black and white defendants will recidivate, but is far more likely to assign a high risk score to black defendants who do not go on to reoffend. White defendants who *did* go on to commit a crime when released were, in turn, more likely to be mislabeled as having low risk. The study established that even when a model is equally accurate in making predictions about members of different racial groups, the false positive and false negative rates might differ between groups. In this case, the costs of false positives (unwarranted incarceration) disproportionately fell on one group.[3]

---

[1] "Artificial Intelligence," *60 Minutes*, October 9, 2016, http://www.cbsnews.com/videos/artificial-intelligence/.
[2] Walker, Joseph. "Meet the New Boss: Big Data: Companies Trade in Hunch-Based Hiring for Computer Modeling," *The Wall Street Journal*, September 20, 2012, http://online.wsj.com/articles/SB10000872396390443890304578006252019616768.

What's more, practitioners seldom provide explanations of the reasons for decisions made by such systems, giving no view of <u>why</u> you got turned down for a job or flagged as a terrorist. As Cathy O'Neil writes in *Weapons of Math Destruction*, "The models being used today are opaque, unregulated, and uncontestable, even when they're wrong."[4] There is some momentum in the research community to fix this,[5] but the issues are complex. There are the reasons why a particular model made a decision—for example, you were denied credit because you've only been at your current job for two months, and you transact with merchants that defaulters frequent. And there are deeper reasons: why are these particular "attributes" deemed by the system to be important evidence of default? Do we want to use this sort of evidence for this sort of decision making? Are the models codifying, and thereby reinforcing, the effects of prior unfairness? Are the models simply incorrect, due to unrecognized biases in the data?  For example, as Kate Crawford and Ryan Calo note, "a 2015 study showed that a machine-learning technique used to predict which hospital patients would develop pneumonia complications worked well in most situations. But it made one serious error: it instructed doctors to send patients with asthma home even though such people are in a high-risk category.[6] Because the hospital automatically sent patients with asthma to intensive care, these people were rarely on the 'required further care' records on which the system was trained."[7]

Year by year, the sophistication of these data-driven algorithms increases, and we already cannot escape their effects. Over the next decade, the data that feed these algorithms will become more pervasive and more personal. Progress toward addressing issues of civil rights and fairness will be made only if incentives are put in place to bypass the considerable roadblocks to success: (1) most computer scientists do not have a deep understanding of issues of fairness and civil rights, and they thus are not traditional or natural issues for computer scientists to address; (2) traditional civil-rights scholars generally lack the sophisticated understanding of big data and data science needed to make substantive progress; (3) the key questions for which answers are needed are not broadly accepted as being important research problems.

<u>Determining if models learned from data exhibit objectionable bias</u>
Establishing whether a model discriminates on the basis of race, gender, age, or other legally protected or otherwise sensitive characteristics might seem like a straightforward task: does the model include any of these features? If not, one might quickly conclude that its decisions cannot exhibit any bias. Unfortunately, there are a number of problems that might nevertheless result in a biased model, even if the model does not consider these features explicitly. Existing scholarly work has addressed the notion that other features can act as surrogates for explicitly sensitive characteristics, most famously location of

---

residence acting as a surrogate for race. However, there are even more complex reasons one might end up with a biased model, despite conscious efforts to the contrary.

The selection of the data used to build the models—the training data—is an important source of potential bias. Non-representative samples of the population will often lead to models that exhibit systematic errors. Such sampling biases are easy to overlook and sometimes impossible to fully recognize; worse, standard validation methods that depend on hold-out data drawn from the same sample will fail to reveal them. Even representative samples—or datasets that capture the entire population of interest—can fail to ensure that models perform equally well for different parts of the population. When minority groups do not follow the same pattern of behavior as the majority group, machine learning may struggle to model the behavior of the minority as effectively as the majority because there will be proportionally fewer examples of the minority behavior from which to learn. Under these conditions, the dominant group in society may well enjoy relatively higher accuracy rates. Training data may also encode prior prejudicial or biased assessments. A model trained on historical hiring data could easily lead to future hiring decisions that simply replicate the discrimination at work in previous human decision-making upon which such modeling hoped to improve. Tainted training examples might wrongly instruct the machine to see features that actually predict success on the job as indicators of poor performance.

Any bias exhibited by such models would be unintentional, but no less pernicious than decisions that explicitly consider legally protected characteristics. Indeed, such models could be *more* pernicious precisely because the bias stems from problems with the training data that are easy to overlook. While data scientists often learn and care deeply about the challenges posed by sampling bias, the difficulty or impossibility of establishing ground truth, and the many ways to measure model performance, there is an urgent need to support research to develop more rigorous methods for establishing whether a model exhibits objectionable bias.

Law and policy often look to disparate impact analyses as a way to establish whether a decision procedure might be discriminatory. Such analyses ask whether the decision-making in question results in a disparity in outcome along lines of race, gender, age or other protected characteristics. If, for example, white job applicants receive offers of employment at a rate 20% higher than black applicants, this might suggest that the process of assessing job applicants suffers from some kind of bias. It might, however, also indicate that the legitimate qualities sought by the employer happen to be held at uneven rates by members of different racial groups. And it can be exceedingly difficult to establish whether a disparate impact stems from the former or the latter—especially when a researcher does not have direct access to the model or training data, or does not fully comprehend the data-generating process.

Much of the research to date that aims to measure bias in systems that rely on machine learning has been performed under adversarial conditions. Outsiders have observed how systems respond to different inputs, and have attempted to uncover ethically salient differences in outputs. Such techniques are commonly known as "algorithmic auditing" and they are what many commentators have in mind when they call for "algorithmic accountability." Researchers operating under these conditions face considerable challenges in making well-justified claims about the source of bias. By necessity, most of these audits have so far focused on systems that are consumer-facing, to which researchers can input some data and observe the output; the result is a series of important cases dominated by instances of machine learning applied to web services. Research is needed not only to explore ways to overcome these challenges, but also to understand whether different methods would be necessary or more effective when organizations attempt to audit their own models or grant outsiders access.

Supporting the emerging field of fairness-aware machine learning
Computer scientists have begun to investigate how concerns with fairness and reducing or eliminating unwanted discrimination might become part of the model-building process. In particular, researchers have developed a number of different formal definitions of fairness that models can be forced to satisfy. One such notion is group parity, which requires that models generate equal outcomes for members of, for example, different racial groups. This requirement, however, might well be in tension with a notion of individual fairness, where assessments are expected to be maximally accurate for each individual. Others have proposed that fairness would be best served by requiring that a machine learning algorithm classify different populations with the same true positive rate; for instance, when such an algorithm is used to decide who gets a loan, the algorithm should have an equal probability of classifying a loan-worthy individual as loan-worthy, irrespective of which subpopulation that individual is from. Still others have worried about cases where accuracy rates are comparable, but where the type of error differs between groups and where these errors have different costs, as in ProPublica's story on recidivism prediction. One group might be subject to a higher rate of false positives (potentially very costly for the affected individual) while the other experiences a higher rate of false negatives (potentially desirable from the individual's perspective).

These approaches have tended to trace the source of unfairness back to different weaknesses in machine learning. Some assume that the main problem resides with training data, which may suffer from all sorts of biases. The task, in such cases, is to compensate for flaws in the data from which the machine will learn. Others have identified cases where machine learning fails to perform as well for minority groups even when the training data is pristine, often because minority groups do not conform to the same patterns as the majority group. This work assumes that the task is to develop methods that can generate models with more even performance across a diverse population, but to do so without having to drag down the model's performance for the majority group.

The field has begun to grapple with the tensions between different notions of fairness, but there is significant disagreement about the appropriate directions for future research. Some see hard trade-offs between competing ideas of fairness; others wonder if data scientists just need incentives to collect more data (both training examples and a larger set of features) to close the gap in performance; still others wonder if tests of validity are even a legitimate measure of fairness, given how deeply bias may suffuse both the training data and the data held out for testing. Shielding machine learning from this taint may require a much more aggressive strategy that assumes that the actual distribution of qualities and capacities across the population is far more equal than compromised data might suggest. Recent scholarship has further complicated this debate by establishing the impossibility of achieving parity across a set of intuitive measures of fairness when groups differ in their underlying rates of important behaviors.[8] Thus hard choices will be necessary, and deep understanding required.

These discussions need to involve a more diverse range of stakeholders, both to improve the quality of the research and to gain legitimacy and buy-in. The values underlying the different notions of fairness need to be debated more openly and explicitly. Computer scientists can contribute in many ways here—e.g., by offering technical solutions, such as taxonomies for fairness and algorithms for achieving fairness. Investment should provide the resources, both material and intellectual, to support and foster these discussions, and to push the field to develop tools that make clear the full range of possibilities for defining and achieving fairness. Future research will also need to consider how organizations would deploy these methods in practice. Paradoxically, most of the methods proposed so far can only achieve fairness by taking class membership (e.g., gender) into account explicitly. In other words, organizations would have to collect information that could easily serve as the basis for intentional discrimination in

---

order to prevent unintentional discrimination. Organizations might also balk at demands to collect more information, even in the interest of improving how well machine learning performs for minority groups, if doing so would appear to intrude on people's privacy, involve significant expense, or create a perceived acknowledgement of wrongdoing or even actual liability.

<u>Providing transparency into and control over the data-driven inferences made about citizens</u>
Discussions and analyses of fairness and bias in algorithmic decision-making are stymied by the difficulty for citizens, users, regulators, and even researchers to understand the reasons that data-driven inferences are made. Research into building so-called "interpretable" or "comprehensible" machine-learned models has received some attention for decades, but many researchers and practitioners still believe that such models—beyond the simplest—are essentially black boxes. Recent work suggests that we can build models that are possible for humans to meaningfully inspect without sacrificing accuracy,[9] but this area needs substantially more attention, both in clarifying what counts as "interpretable" or "comprehensible" and when such properties are desirable or necessary.[10]

However, building such models is only one strand of research into transparency of data-driven inferences. Often the crucial interest is not in understanding the model, but in understanding the precise reasons for a particular inference. As discussed above: I was denied credit. Why? Research into explaining the decisions made by data-driven systems is even sparser than research into comprehensible models—but possibly more practically useful. Encouragingly, explaining an individual decision may actually be easier than explaining a complex model behind the decision.[11] For example, one may take a counterfactual approach:[12] considering the algorithm input as a collection of evidence, what is the minimal set of evidence the removal of which would inhibit the inference? If we were interested in providing an explanation that might inform future behavior, we might instead ask: which features (characteristics, aspects of behavior, etc.) would be the least costly for an individual to change, so as to produce the desired change in the inference?[13] We need much more research on explaining inferences, and doing so efficiently and effectively, before we can be confident that we are indeed giving sufficient transparency.

As a society, we also may want to give citizens control over the data that are used to make inferences about them. Someone may not want their visit history to gay rights websites to be used in decisions—automatic or otherwise—that are made about him or her. Decision-specific transparency seems to be a prerequisite for giving such control.[14]

More deeply, in order truly to understand the civil rights implications of data-driven systems, we don't only need to understand the models and the decisions that they make, but we also need to understand why the models are as they are! This ties model comprehensibility to all the other research streams described in this document, as models are as they are because of the selection of machine learning algorithm, the selection of training data (and evaluation data), and more insidiously, many other decisions made in the process of formulating the problem and the evolution of the system.

<u>Looking beyond the algorithm for the sources of unfairness, discrimination, etc.</u>

---

[9] Zeng, Jiaming, Berk Ustun, and Cynthia Rudin, "Interpretable classification models for recidivism prediction," arXiv:1503.07810, 2015.

[10] Lipton, Zachary C, "The mythos of model interpretability," arXiv:1606.03490, 2016.

[11] Martens, David, and Foster Provost, "Explaining Data-Driven Document Classifications," *MIS Quarterly* 38.1 , pp 73-99, 2014.

[12] Ibid. and Chen, et al. "Enhancing Transparency and Control when Drawing Data-Driven Inferences about Individuals," 2016 ICML Workshop on Human Interpretability in Machine Learning, https://arxiv.org/abs/1606.08063.

[13] NB: This is a much more difficult problem than the former, as it requires modeling causal relations, costs/benefits, and statistical dependencies involving the individual, rather than just causal relations between the input and output of the model.

[14] Ibid. and Chen, et al. "Enhancing Transparency and Control when Drawing Data-Driven Inferences about Individuals," 2016 ICML Workshop on Human Interpretability in Machine Learning, https://arxiv.org/abs/1606.08063.

One very crucial aspect of data-driven decision-making is only just beginning to be taken into account in discussions of and research into ethics, data science, and civil rights, although it does not surprise savvy practitioners: the technical formulation of the problem makes all the difference. It has long been accepted within data science circles that building data-driven systems is a process[15] that involves carefully understanding the problem to be addressed, understanding the data available (sometimes at a cost), and formulating the problem to which machine learning/statistical inference algorithms will be applied. Take as an example supervised predictive modeling. Formulating the problem involves: deciding on the instances to be modeled, crafting a definition of the target variable, obtaining "labels" (ground truth or proxies for it) for the training data, selecting an appropriate sample of the data from which to train, and then engineering a set of features that will be predictive (or using algorithms that build the features autonomously). In practice, each of these choices often incorporates approximations, proxies, surrogates, and biases. In the ideal case, these are chosen with full consideration of the likely (side) effects, but in practice, not only are the consequences of the choices not apparent; sometimes the biases and approximations are not even well understood by the researchers and practitioners. For example, an unseen racial bias in prior decision-making may be recapitulated in models learned from those data. An unrecognized selection bias in data sampling may miss a critical subpopulation—for example, consider sampling data from users of smartphones.

A robust understanding of the ethical use of data-driven systems needs substantial focus on the possible threats to civil rights that may result from the formulation of the problem. Such threats are insidious, because problem formulation is iterative. Many decisions are made early and quickly, before there is any notion that the effort will lead to a successful system, and only rarely are prior problem-formulation decisions revisited with a critical eye. In addition, systems whose underlying knowledge evolves over time, be it via continual machine learning or manual intervention, may themselves be making implicit decisions on (for example) the selection of the population. What are the implications for fair treatment, if fair treatment is not considered in the design of the selection mechanisms? A system that starts with a small bias may unwittingly magnify it.

Creating cross-disciplinary scholars
The recognition that applications of big data and data science can implicate civil rights has spurred calls for greater involvement of social scientists, lawyers, and policy experts in the development, deployment, and review of data-driven systems. While laudable, calls for cross-disciplinary collaboration, and especially collaborative research, on these issues rarely consider the challenges that members from these different communities will face when attempting to engage with one another. Those who work with data science, including computer scientists, are rarely trained in law and policy; experts in civil rights rarely have a background in machine learning or computational statistics. Neither is well prepared to identify the many ways that a particular application of data science may implicate civil rights—or what to do in such cases.

The difficulty of cross-disciplinary collaboration has already led to a rather troubling pattern in work on civil rights and data science. Critical writing often struggles to recognize how machine-learned systems differ from other types of formalized decision-making and how these differences introduce novel dangers for civil rights, only some of which can be effectively addressed with standard policy instruments. Likewise, data science researchers concerned with civil rights have tended to tackle issues of fairness as if such weighty topics have not already received considerable attention in law, social science, and philosophy.

Work integrating civil rights and data science cannot be easily divided between collaborators, where the more expert team member would handle each task. Meaningful legal analysis will require technical expertise; rigorous technical review will depend on a nuanced understanding of legal concepts. Valuable breakthroughs are most likely to come from researchers who combine expertise in both domains. Future investment in research should foster collaborations that do more than put different communities in contact; investment should support the training necessary to cultivate a future generation of researchers who are simultaneously expert in both fields.

While facilitation of collaboration between researchers is important, a workforce that understands the relationships between ethics and data is also key. The recent groundswell of interest in data science, and the emergence of new interdisciplinary graduate and undergraduate programs in this area, provide a natural opportunity here.  The challenge is that these curricula are already overly crowded and often patched together from existing courses in different departments. Adding a single course about data, ethics, algorithmic bias, fairness, accountability and the law to such a program would be a good start; a better idea would be to thread the associated ideas through all of the courses in a coherent manner. As data science methods and tools become integral and essential elements of research across a broad range of disciplines, it would be worthwhile to broaden the existing research ethics training requirements to include these aspects. A national-level conversation about those important ideas—what they are and how to teach them—could help individual institutions with those initiatives.

Needless to say, it will be important to engage the educational research community in that conversation. Data science is also making inroads into the high-school curriculum, which provides even earlier opportunities to weave together knowledge about ethics, law, and data. There the issues are somewhat different; not only crowded curricula, but also the forces of standardized testing, as well as teacher training. Development of easy-to-deploy materials that convey the important concepts and issues, while also aligning with existing learning objectives, will be essential to success here. This, too, should involve the educational research community. Initiatives like the National Science Foundation's "STEM + C"[16] and the National Academy of Engineering's series of workshops in this area[17] can also usefully inform these conversations.

Several recent meetings and reports have brought to light issues and concerns regarding big data, data science, algorithmic decision-making, and civil rights:

- The White House issued Big Data reports in 2014, notably raising concerns about discrimination and the inscrutability of algorithms and calling for more research to establish how data science might implicate civil rights and how to mitigate against these dangers.
- The Federal Trade Commission issued its own report at the start of 2016, laying out how existing laws apply to commercial use of data science that raise concerns with discrimination and fairness, but also noting gaps in policy where commercial actors should exercise careful judgment, despite the lack of clear technical or ethical guidance.
- The White House followed with a more detailed and narrowly focused report in 2016 on "Big Data: A Report on Algorithmic Systems, Opportunity, and Civil Rights", which considered how data science could be a boon—but also a threat—to civil rights in the areas of consumer credit, employment, education, and criminal justice.
- Alongside this report, the White House also released a document outlining a National Privacy Research Strategy, which, notably, called for new work on the dangers posed by "analytical algorithms" as issues distinct from traditional privacy concerns.
- The White House recently issued another report on artificial intelligence that laid out in greater technical detail concerns with "Fairness, Safety, and Governance"—and the need for further investment in and research on these topics.

*This material is based upon work supported by the National Science Foundation under Grant No. 1136993. Any opinions, findings, and conclusions or recommendations expressed in this material are those of the authors and do not necessarily reflect the views of the National Science Foundation.*